# Simulation of atomic structure in the neighbourhood of nanovoids in bcc Fe

A.V. Nazarov[1,2], I.V. Ershova[1] and Y.S. Volodin[1]

[1]National Research Nuclear University MEPhI (Moscow Engineering Physics Institute) Kashirskoe highway 31, Moscow,115409, Russia
[2]Alikhanov Institute for Theoretical and Experimental Physics (ITEP) Russia

**E-mail**: AVN46@mail.ru

**Abstract**. Generally displacement fields in the vicinity of voids were determined by the equations of theory of elasticity. Such a description has its disadvantages as it doesn't take into account the discrete atomic structure of materials and it should be expected that atom displacements in the vicinity of nanovoids should significantly differ from the predictions of mentioned theory. In our recent works a new variant of Molecular Static method was developed. In particular in this model an iterative procedure is used in which the atomic structure in the vicinity of defect and constant, determining the displacement of atoms embedded into an elastic continuum, are obtained in a self-consistent manner. In this work we use our model for investigation of the atomic structure in the vicinity of nanovoids of different sizes. Results show that the displacements are significantly different for variant crystallographic directions and these differences are particularly large in bcc metals.

## 1. Introduction

The voids play the significant part in the processes of material structure forming, diffusion phase transformations, swelling, etc. Therefore it is necessary to develop the methods of determining the defects characteristics. Also it's obvious that defect characteristics are determined by the atomic structure. Atoms surrounding defect shift from the sites of ideal lattice, e.g. defect atomic structure changes with respect to an ideal one, that in turn leads to changes in interaction energy of neighbor atoms and results in modification of defect energy characteristics and other features. Generally displacement fields in the vicinity of point defects as well as nanovoids were determined by the solution of equations from the classical theory of elasticity [1], were displacement field has a view:

$$u_{x_i} = C_0 x_i + C_1 \frac{x_i}{r^3}, \qquad C_0 = \frac{1-2\nu}{E}\frac{2\gamma R^2}{R_G^3 - R^3}, \qquad C_1 = -\frac{1+\nu}{E}\frac{\gamma R^2 R_G^3}{R_G^3 - R^3}, \qquad (1)$$

were $r$ is a distance from a void center, $x_i$ are coordinates, $R$ is a radius of void, $R_G$ is a radius of district containing the void, $\nu$ is the Poisson ratio and $E$ is the Young modulus, $\gamma$ is the surface energy. Such a description has its disadvantages as it doesn't take into account the discrete atomic structure of materials. Results of classical elasticity theory are expected to be valid at distances from a defect that are much larger than the lattice parameter which is a characteristic feature of discreteness, therefore the quantities of atom displacements in the vicinity of such defects as vacancies, vacancy complexes and nanovoids should significantly differ from the predictions of these displacements obtained by means of theory of elasticity. In our recent works a new approach was developed [2-4]. In particular in this approach an iterative procedure was used in which the atomic structure in the vicinity of point defect and constant, determining the displacement of atoms embedded into an elastic continuum with accordance with asymptotic solution of equations from the classical theory of elasticity, are obtained in a self-consistent manner. The vacancy features (including formation volumes and migration volumes) obtained for a number of cubic metals agreed well with experimental values [3,4]. We also note that the MD simulation results, that are concerned the self-diffusion in bcc iron under pressure [5], agree within error of MD experiment with the data of our work [3]. We used our model for simulations of the di-vacancy features [6] and later for simulations of the vacancy complexes [7]. In this work we use our approach for direct investigation of the atomic structure in the vicinity of nanovoids in some bcc metals.

## 2. Model for point defect

Equilibrium positions of atoms in computation cell are simulated by using a variational procedure analogous to that is usually used in Molecular Static Method [8]. Computation cell is rounded by atoms embedded in an elastic continuum and displacements of these atoms $\boldsymbol{u}$ concerned with perturbations, which are induced by point defect, are calculated using solutions of static isotropic elastic equation [1]:

$$(\lambda + 2\mu)\nabla(\nabla \cdot \boldsymbol{u}) - \mu\nabla \times (\nabla \times \boldsymbol{u}) = 0, \qquad (2)$$

where $\lambda$ is Lame modulus, $\mu$ is share modulus. Solution of this equation can be expressed in term of spherical harmonics and thus there are an infinite number of solutions to displacement of each atom $\boldsymbol{u}$ as a function of its position $\boldsymbol{r}$ (i.e. function of its distance from a defect). We use the first two terms of series for atomic displacements calculation in an elastic matrix. First term has spherical symmetry and can be written like Eq. 1 and has only first term when $R_G \gg R$ [7]:

$$\boldsymbol{u}_1 = C_1 \boldsymbol{r}/r^3. \qquad (3)$$

Next term of series has cubic symmetry and usually written in the form [8]:

$$\boldsymbol{u}_2 = C_2 \nabla \left[ \frac{1}{r^5} \left( \frac{x^4 + y^4 + z^4}{r^4} - \frac{3}{5} \right) \right], \qquad (4)$$

where $C_2$ is a constant.

Prior calculations carried out for vacancy in bcc iron using Johnson's potential [8] have shown that atomic displacements calculated taking into account only first spherically symmetric term of series don't agree with results of variation computations even for sufficiently large systems. This circumstance becomes apparent when the formation and migration volumes are estimated. Taking into account second term of series (Eq. 4) allows to define atomic displacements in an elastic matrix more precise and consequently equilibrium positions of all atoms in the vicinity of a defect are also defined more precise. Other terms of series have symmetry, which is differ from atomic structure symmetry in the vicinity of a defect, or they very fast decrease with $\boldsymbol{r}$ and their contributions in $\boldsymbol{u}$ are negligible at distances corresponding to atom locations in an elastic continuum. Self-consistent iteration procedure for calculation constants $C_1$, $C_2$ and simulation atomic structure in defect crystal is realized in our model. Stable convergence upon constants $C_1$ and $C_2$ has been received. This procedure and our model in all have been described in detail [3,4].

## 3. Results and Discussion

The simulation is done for voids of different sizes. We used N-body potentials developed in [9] for BCC iron. The results of simulation are presented on Figure 1 and also in Table 1. In particular, there give typical atom displacement dependencies on the distance from the void center and the calculation results on the equations for the single isolated void (Eq. 1 and Eq. 3). Dashed vertical lines indicate the radiuses of the void and the computational cell respectively.

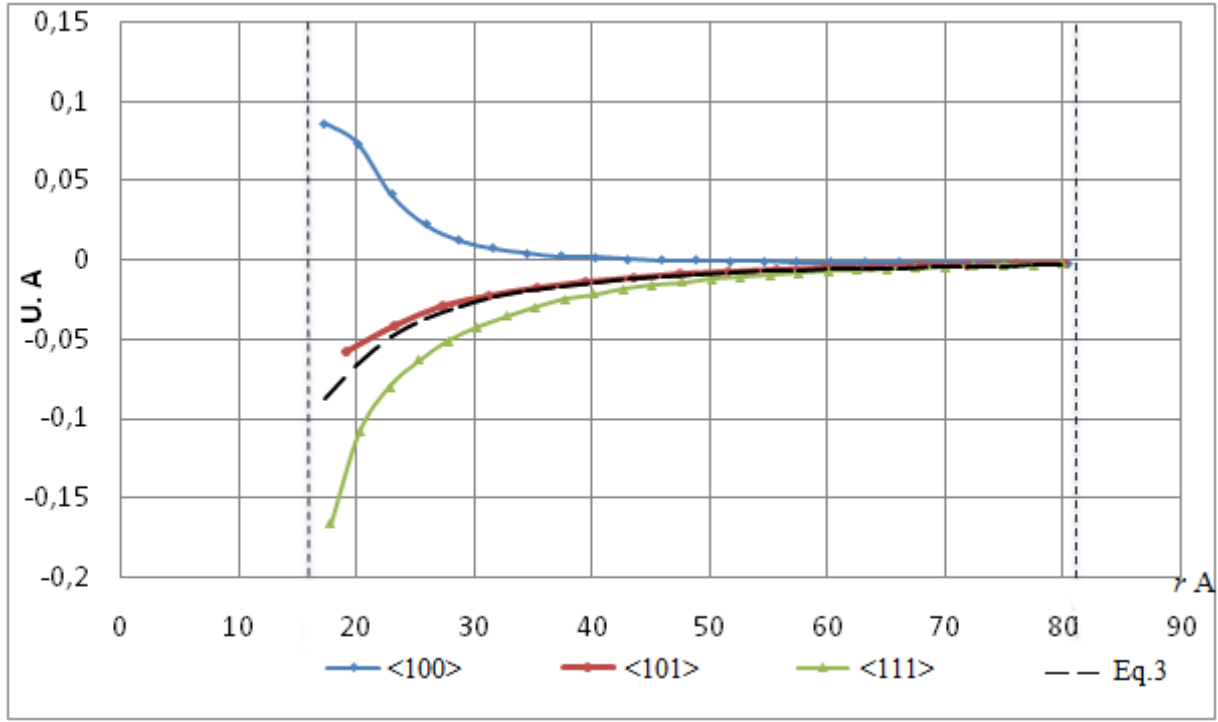

**Figure 1**. BCC Fe. Displacement field $U$ for nanovoid ($R$ =17.46 Å)

Table 1. Simulation Results. BCC Fe

| R Å | 8.72 | 10.49 | 12.17 | 15.37 | 17.46 | 20.06 |
|---|---|---|---|---|---|---|
| U<100> Å | 0.029 | 0.034 | 0.050 | 0.066 | 0.086 | 0.091 |
| U<101> Å | -0.021 | -0.026 | -0.031 | -0.040 | -0.057 | -0.044 |
| U<111> Å | -0.083 | -0.081 | -0.100 | -0.113 | -0.166 | -0.170 |
| $C_1$ Simulation | -1.601 | -3.763 | -4.255 | -8.915 | -13.666 | -28.635 |
| $C_1$ Theory | -4.84 | -7.00 | -9.42 | -15.03 | -19.4 | -25.61 |

Results show that the fields of displacements in the vicinity of nanovoids are significantly more complicated and with much bigger magnitudes of displacements than in the vicinity of vacancies [2-4]. In addition, the displacement significantly different for variant crystallographic directions, and these differences are particularly large in metals with bcc structure. The highest value of the atom displacements near the void surface increases with increasing void size (Table 1). It should be emphasized that the positive displacement of the atoms located near the surface in the <100> direction increase with the size of the voids, instead of decrease as one would expect. In addition, it is interesting that the trend is continued, although the void radius value exceeds 20 A in the simulation (Table 1). Thus, discreteness of structures plays an important role in the formation of the atom displacements near the void surface.

It should be noted an important consequence of the results. Usually, the equation of vacancy diffusion in the presence of stress field has the following form [10]:

$$\vec{J} = -D_V \left[ \nabla c + c \frac{V_{Rel}}{kT} \nabla (\mathrm{Sp}\sigma) \right], \tag{5}$$

were $c$ is the vacancy concentration, $\sigma$ is a tensor of stress, $V_{Rel}$ is a relaxation volume.

A similar expression we have obtained in the zero approximation in the framework of the microscopic approach proposed by us to describe the effect of the elastic fields on the diffusion fluxes of vacancies [11]:

$$J_x = -\frac{1}{\Omega} \left[ D_V \frac{\partial c}{\partial x} - c \frac{K^V}{kT} \left( D_V \frac{\partial \mathrm{Sp}\varepsilon}{\partial x} \right) \right], \tag{6}$$

where $\Omega$ is the volume per lattice site, $D_V$ is the vacancy diffusion coefficient in the perfect system and $\varepsilon_{ij} = (1/2)(\partial u_i/\partial x_j + \partial u_j/\partial x_i)$ is the strain tensor ($i, j$ = 1,2,3).

$$K^V = \frac{1}{2} \sum_s \sum_{k \neq s} \frac{(x_{ks}^V)^2}{R_{ks}^V} \frac{\partial E_{Sys}}{\partial R_{ks}} \bigg|_{R_{ks}^V} \tag{7}$$

$x_k, y_k, z_k$ are the coordinates of the atom $k$ of a system, $E_{Sys}$ is the system energy, $x_{ks} = x_k - x_s, y_{ks} = y_k - y_s, z_{ks} = z_k - z_s, k \neq s, R_{ks} = |\mathbf{r}_k - \mathbf{r}_s| = \sqrt{x_{ks}^2 + y_{ks}^2 + z_{ks}^2}$
for all atoms, $x_{ks}^V, y_{ks}^V, z_{ks}^V, R_{ks}^V$ are the coordinate differences between the atoms $k$ and $s$ and spacing's between them in system with the vacancy.

It is known that the vacancy flux density on the void surface determines growth rate of the voids [12]. If we use the solution of equations from the classical theory of elasticity in the vicinity of the void (1), then:

$$\mathrm{Sp}\varepsilon = 3C_0, \quad \nabla \mathrm{Sp}\,\varepsilon = 0, \quad \text{and} \quad \vec{J} = -D_V \nabla c. \tag{8}$$

There is not stress influence on flux of vacancies [12,13].

Obtained results show that $\nabla \mathrm{Sp}\varepsilon \neq 0,$ and a kinetics equation for the growth rate of voids must contain the additional terms conditioned by strains, arising from voids. Thus, these factors, together with the factors which change the diffusion coefficients [13], affect the kinetics of pore growth under different conditions. It should be take into account that in many cases the growth and dissolution of pores pass through the stage of nano-sizes.

One more consequence: it is to be expected that the displacement fields around nano-sized precipitates will also significantly differ from the predictions of the elasticity theory. Therefore, the existing theories and models for nucleation and growth of pores and phases' inclusions need a significant revision, taking into account the results of the simulation. The above-mentioned consequences are especially relevant for the swelling models of materials [13] and Frenkel's effects (Kirkendal of the second kind) [14, 15] with interdiffusion.

## 3. Conclusion

New model is presented for determining atomic structure in the vicinity of nanovoids. The substantial points of this model are as follows. First, we allow for displacements of atoms embedded in an elastic continuum around computation cell. Second, we consider influence of discrete nature of the atomic structure in the vicinity of a defect on values and directions of atomic displacements in elastic matrix.

The third point is that the atomic structure in the vicinity of the void and the displacement of atoms in the elastic medium are calculated in a self-consistent manner using a convergent iterative procedure.

The model permit to find qualitatively new peculiarities of the atomic structure in the void vicinity that cannot be correctly obtained with the use of the elasticity theory because it cannot be applied to the atomic scale.

In addition, the displacement significantly different for variant crystallographic directions.

The obtained atoms positions give us the possibility to provide the following more advanced level simulation of nanovoid growth in materials supersaturated with vacancies.

**Acknowledgements**

We wish to thank student Ustin Safonov for help in some calculations.